\documentclass[11pt]{article}

\usepackage{epsf}
\usepackage{epsfig}
\usepackage{cite}
\usepackage{amsmath}
\usepackage{graphics}
\usepackage{xcolor}

\begin{document}
\begin{titlepage}

\hfill Revised

\hfill February 2024\\

\begin{center}

{\bf \LARGE Kerr-Newman and Electromagnetic Acceleration}\\
\vspace{2.5cm}
{\bf Paul H. Frampton}\footnote{paul.h.frampton@gmail.com}\\
\vspace{0.5cm}
{\it Dipartimento di Matematica e Fisica "Ennio De Giorgi",\\ 
Universit\`{a} del Salento and INFN-Lecce,\\ Via Arnesano, 73100 Lecce, Italy.
}

\vspace{1.0in}

\begin{abstract}
\noindent
Previous discussions of charged dark matter neglected Primordial Black Hole (PBH) spin
and employed the Reissner-Nordstrom metric. In Nature we expect
the PBHs to possess spin which require use of the technically more challenging
Kerr-Newman metric. It is shown that the use of K-N metric retains the
principal properties already obtained using the R-N metric, in particular
the dominance of Coulomb repulsion requires super-extremality
and the presence of naked singularities. In this sense, the spin of the PBHs
is not an essential complication. 
\end{abstract} 
\end{center}

\end{titlepage}

\noindent
\section{Introduction}

\noindent
In \cite{FramptonPLB,FramptonMPLA}, a cosmological model
was introduced which departs from the conventional wisdom that the energy
pie for the visible universe is approximately 5\% normal matter,
25\% dark matter and 70\% dark energy. The term dark energy was
introduced in 1998 in association with the then newly-discovered 
accelerated expansion. In the novel model, the energy pie is revised
to 5\% normal matter, 95\% dark matter and 0\% dark energy shich is
replaced by  what we may call charged dark matter.

\bigskip

\noindent
We are discussing a toy model which is not intended to describe Nature
precisely but is intended to be sufficiently realistic that some general suggestions
about the origins of dark energy can be made. The dark matter constituents
are assumed to be PBHs with a spread of masses from  one hundred
to at least one trillion solar masses.

\bigskip

\noindent
We shall not endeavour to invent a constituent of dark energy directly but
shall conclude that dark energy, describable in an opaque manner
by a cosmological constant
$\Lambda\sim (meV)^4$, and also possibly interpretable as a vacuum energy,
can be better understood as a result of
electromagnetic forces between extremely heavy PBHs which are
a sector of dark matter but, unlike lighter PBHs, carry electric charge.

\bigskip

\noindent
The present article is concerned with one issue which is whether the spin
of the PBHs is an essential complication. To check this out, we must promote
the Riessner-Nordstrom metric used in \cite{FramptonMPLA} to the
Kerr-Newman metric which is technically more complicated and, again
to anticipate our result, we shall conclude that spin does not alter our
main conclusions that the PBHs need to be super-extremal and to possess
naked singularities.

\bigskip

\noindent
This maximally violates the cosmic censorship hypothesis, which is
nevertheless unproven.

\bigskip

\noindent
The biggest assumption in this approach to replace dark energy by
charged dark matter is that Primordial Extremely Massive Black Holes (PEMBHs) with unscreened electric charges
exist. It is normally expected that electric charges are screened from
distant observers by accretion of material with counterbalancing charge.
While this is undoubtedly true for stars, planets and galaxies,there is no observational
evidence for or against this compensation of charge happening for this
heavier mass range.

\bigskip

\noindent
Before PEMBHs are discovered, a promising precursor would be the
discovery of Primordlal Intermediate Mass Black Holes (PIMBHs) in the mass range 100 to $10^5$ solar masses
as constituents of the Milky Way dark matter. This may well be possible 
in the foreseeable future using
microlensing of the stars in the Magellanic Clouds by, for example, the
Vera Rubin Observatory (formally LSST) in Chile.

\newpage

\section{Kerr-Newman metric}

\noindent
The most general metric for describing a black hole with non-zero
charge and spin is \cite{Kerr}
\begin{equation}
ds^2 = c^2d\tau^2 = - \left(\frac{dr^2}{\Delta} + d\theta^2 \right) \rho^2
+ \left(cdt - a\sin^2\theta d\phi \right)^2 \frac{\Delta}{\rho^2}
-\left( (r^2+a^2)  d\phi - ac dt \right)^2  \frac{\sin^2 \theta}{\rho^2}
\label{KNmetric}
\end{equation}
in which $(r, \theta, \phi)$ are spherical polar coordinates and

\begin{equation}
a = \frac{J}{Mc}; ~~~ \rho^2= r^2 + a^2 \cos^2\theta; ~~~ \Delta= r^2 - r_Sr +a^2 + r_Q^2 = r^2fr)
\end{equation}
with $f(r) =\left( 1 - r_s/r + r_Q^2/r^2 \right)$ and
\begin{equation}
r_S = \frac{2 G M}{c^2}; ~~~ r_Q^2 = \frac{Q^2 G}{4\pi \epsilon_0 c^4}
\end{equation}

\bigskip

\noindent
The coefficient of $dr^2$ is

\begin{equation}
-\left( \frac{\rho^2}{\Delta} \right) = -  \frac{r^2 + a^2 \cos\theta}{r^2 f(r)}. 
\label{dr2}
\end{equation}

\bigskip

\noindent
To obtain the gravitational stress-energy tensor $T_{\mu\nu}^{(GRAV)}$ involves taking the second derivative
of the metric by employing the expression

\begin{eqnarray}
T_{\mu\nu}^{(GRAV)} &=& -\frac{1}{8\pi G} \left( G_{\mu\nu} + \Lambda g_{\mu\nu} \right). \nonumber \\
& & + \frac{1}{16 \pi G (-g)} \left( (-g)(g_{\mu\nu}g_{\alpha\beta}-g_{\mu\alpha}g_{\nu\beta}) \right)_{,\alpha\beta}
\label{TGRAV}
\end{eqnarray}
where the final subscripts represent simple partial derivatives. We shall need also to evaluate the electromagnetic
counterpart

\begin{equation}
T_{\mu\nu}^{(EM)}  =  F_{\mu\alpha} g^{\alpha\beta} F_{\mu\beta} 
- \frac{1}{4} g_{\mu\nu} F^{\alpha\beta}F_{\alpha\beta}.
\label{TEM}
\end{equation}

\bigskip

\noindent
Let us begin with $T_{\mu\nu}^{(GRAV)}$. This calculation involves taking up to
the second derivatives of the KN metric in Eq.(\ref{KNmetric}). For the function $f(r)$:

\begin{equation}
\frac{\partial}{\partial r} f(r) \sim O(1/r^2)  ~~~~~ \frac{\partial^2}{\partial^2 r} f(r) \sim O(1/r^4)
\label{f1}
\end{equation}
\begin{equation}
\frac{\partial}{\partial r} (f(r))^2 \sim O(1/r^2)  ~~~~~ \frac{\partial^2}{\partial^2 r} (f(r))^2 \sim O(1/r^4)
\label{f2}
\end{equation}
\begin{equation}
\frac{\partial}{\partial r} (f(r))^{-1} \sim O(1/r^2)  ~~~~~ \frac{\partial^2}{\partial^2 r} (f(r))^{-1} \sim O(1/r^4)
\label{f3}
\end{equation}
\begin{equation}
\frac{\partial}{\partial r} (f(r))^{-2} \sim O(1/r^2)  ~~~~~ \frac{\partial^2}{\partial^2 r} (f(r))^{-2} \sim O(1/r^4)
\label{f4}
\end{equation}

\bigskip

\noindent
It is not difficult to check, using these derivatives, that the 1st, 3rd and 4th terms
in Eq.(\ref{TGRAV}) all fall off as $O(1/r^2)$ and that only the 2nd term does not.

\bigskip

\noindent
In $T_{\mu\nu}^{(EM)}$,
we can straightforwardly see that both terms in Eq.(\ref{TEM}) fall off as $O(1/r^2)$.

\bigskip

\noindent
We deduce that
\begin{equation}
T_{\mu\nu}^{(GRAV)} + T_{\mu\nu}^{(EM)} = - \left( \frac{\Lambda}{8\pi G} \right) g_{\mu\nu} + O(1/r^2).
\label{Ttotal}
\end{equation}

\bigskip

\noindent
Thus, the principal conclusions for spinless PBHs found in \cite{FramptonMPLA} carry over to PBHs with spin. In particular, the 1/r expansion is unchanged and the EoS  for the dark energy remains
$\omega = -1$.  The requirement for super-extremality and naked singularities, to allow
Coulomb repulsion to exceed gravitational attraction, remains the same as in the spinless case.

\newpage

\section{Varying cosmological "constant'' $\Lambda(t)$}

\noindent
The Freidmann expansion equation keeping only dark energy is
\begin{equation}
\left( \frac{\dot{a}}{a} \right)^2 = \frac{\Lambda_0}{3}
\label{FriedmannLambdaZero}
\end{equation}
which, normalizing to $a(t_0)=1$, leads to a time dependence
of $a(t)$ which is exponentially growing 
\begin{equation}
a(t) = \exp \left( \sqrt{\frac{\Lambda_0}{3}} (t-t_0) \right)
\label{exponential}
\end{equation}

\bigskip

\noindent
In the EAU-model, the dark energy is replaced by charged dark matter
and we know from \cite{FramptonMPLA} that the equation of state for
the dark energy is precisely $\omega = -1$. 

\bigskip

\noindent
However, the time-dependence of $a(t)$ is no longer exponential. If we use
the dual dark energy description it is governed not by Eq.(\ref{FriedmannLambdaZero})
but by a time-dependent cosmological "constant" $\Lambda(t)$
\begin{equation}
\left( \frac{\dot{a}}{a} \right)^2 \approx \frac{\Lambda(t)}{3}
\label{FriedmanCDM}
\end{equation}

\bigskip

\noindent
The time-dependence follows from the usual dilution of dark matter
\begin{equation}
\Lambda(t) \approx  a(t)^{-3} \approx t^{-2}
\label{timedep}
\end{equation}
so that adopting as the present value $\Lambda(t_0) = (2 ~ meV)^4$
we arrive at the examples of future values for $\Lambda(t)$ provided in Table 1.
In the last row, for example, at a time which is $10^5$ times the present age
of the universe,
the cosmological constant has decreased by a factor $\sim 10^{-9}$.  This
behaviour implies that the extroverse never expands to the very large size
discussed for the $\Lambda CDM$-model in \cite{FramptonIJMPA}.

\begin{table}[t]
\caption{COSMOLOGICAL ``CONSTANT".}
\begin{center}
\begin{tabular}{||c|c||}
\hline
\hline
time & $\Lambda(t)$ \\
\hline
\hline
$t_0$ & $(2.0meV)^4$ \\
\hline
\hline
$t_0+10Gy$ & $(1.0 meV)^4$ \\
\hline
\hline
$t_0+100Gy$ & $(700 \mu eV)^4$ \\
\hline
\hline
$t_0+1Ty$ & $(230 \mu eV)^4$ \\
\hline
\hline
$t_0+1Py$ & $(7.4\mu eV)^4$ \\
\hline
\hline
\end{tabular}
\end{center}
\end{table}

\bigskip
\bigskip
\bigskip

\newpage

\section{Previous Work}

\noindent
There are in the literature, several papers on theoretical
astrophysics and cosmology that cast possible doubt on whether
electrically-charged black holes exist in Nature and 
on whether small charge asymmetries can exist within galaxies
and clusters of galaxies, and within the whole visible universe.
Since the existence of charged black holes is crucial to the EAU model
\cite{FramptonPLB,FramptonMPLA}, the present section addresses
past work and attempts to convince the reader that there is
no known fatal flaw in EAU.

\bigskip

\noindent
It is generally accepted that electrically neutral black holes
exist in Nature. Since the main result of the present article
is that black hole spin is an inessential complication in the
EAU model, we may associate electrically neutral BHs with
the Schwartzschild  ({\underline{S}) metric and electrically
charged BHs with the Reissner-Nordstrom (\underline {RN})
metric. Both \underline{S} and \underline{RN} metrics are exact solutions of general
relativity. The \underline{RN} metric is richer than the \underline{S} metric in that,
depending on the electric charge Q, relative to the mass M,
it has sub-extremal (two horizons), extremal (one horizon)
and super-extremal (no horizon) varieties, Mathematically,
however, both \underline{S} and \underline{RN} are solutions of Einstein's field
equations and there appears to be
no special reason why Nature should simultaneously adopt the \underline{S} solution
and eschew the \underline{RN} solution.

\bigskip

\noindent
It is of considerable interest that in 1959 Lyttleton and Bondi
\cite{LyttletonBondi} considered electric
charge as underlying the cosmological expansion, by introducing a small
difference between the proton and electron charges. Their theory did not
survive quantitative testing of the neutrality of the hydrogen atom. The LB
model is obviously
different from the EAU model, although it does involve a remarkably
similar charge asymmetry $\epsilon_Q \sim 10^{-18}$ for the universe
as we have for clusters.

\bigskip

\noindent  By studying the
baryon budget\cite{Rasera,Durier} and making reasonable
assumptions, one can show that at a scale of
2Mpc in the EAU model that, unlike in the LB model, there is no charge asymmetry. 
This scale is larger than a galaxy. In the EAU model, all processes conserve
electric charge and so certainly the universe remains electrically neutral unless 
there was, for an unknown reason, an initial condition that imposed
$|\epsilon_Q| \neq 0$ at the Big Bang.

 \bigskip
 
 \noindent
 One study of the cross-sections for capture of particles by an
 RN black holes was made by Zakharov\cite{Zakharov}. This
 study included, however, capture only of electrically neutral particles and
 hence neglected the effects of Coulomb forces,

\bigskip

\noindent
In the EAU model, we have assumed that the fraction $f_{DM}$ of the dark
matter made up of PBHs is $f_{DM} =1$. The PBHs fall into different
mass ranges for PIMBHs, PSMBHs and PEMBHs and in the example
we gave the PEMBHs contribute $f_{DM} = 0.5$.  In \cite{CarrSilk}
an upper limit $f_{DM}=0.01$ for mass $M=10^{12}M_{\odot}$
which is contradictory. However, we have reason to believe
that the constraints in \cite{CarrSilk} are far too severe. This can be
traced back to the well known ROM paper\cite{ROM} which introduced
the popular $f_{DM}$ exclusion plot. 

\bigskip

\noindent
In the case of ROM, too severe limits were caused by the use,
in numerical analysis, of an oversimplified accretion
model (Bondi model) which assumes spherical symmetry and radial inflow. This
naive model can overestimate accretion, compared to Nature, not just by a factor 2
but even by some 4 or 5 orders of magnitude!\cite{M87}. Subsequent exclusion plots 
appear to have been overly influenced by ROM's
results. The fact that ROM imposed too severe constraints has been
confirmed to us by its senior author (J.P. Ostriker, private
communication).

\bigskip

\noindent
 There are other mechanisms proposed in the literature able to produce 
 a charged BH. For instance, in \cite{Wald}, the author argued that a rotating 
 BH embedded in a plasma may generate a magnetic field and may acquire 
 an electric charge. However, a recent investigation in \cite{Komissarov}
 has shown that the charge in \cite{Wald} is in fact an upper limit and 
 generally is quite small.
 
 \bigskip
 
 \noindent
 To summarise this section, previous work on these issues is
 undoubtedly interesting but does not reveal a fatal flaw
 in the EAU model which replaces dark energy by charged
 dark matter.
 
 \bigskip
 
 \noindent
 Dark energy can be simply and accurately parametrised by the cosmological constant
 $\Lambda$ introduced
 by Einstein for a different, almost antithetical, reason. It
 may equally be regarded as a type of vacuum energy. In particle
 theory, however, vacuum energy is associated with the spontaneous breaking of a
symmetry and it is unclear which symmetry is involved.

\bigskip

\noindent
The EAU model can provide a more physical picture than either. The model is
very speculative and cries out for more theoretical work to confirm
its internal consistency.

\newpage

\section{Discussion}

Dark matter and dark energy are the two major unanswered questions
in cosmology. In this article we have discussed a model
which suggests specific answers. The model has the following ingredients.

\bigskip

\noindent
In the Milky Way, the dark matter constituents are PIMBHs with masses between
100 and 100,000 solar masses. We cannot exaggerate the importance of
detecting these, in our opinion most easily by seeking microlensing light curves
with durations in excess of two years. Such a success would suggest that the
idea that DM constituents are PBHs is on the right track.

\bigskip

\noindent
PBHs with higher masses between a million and a hundred billion solar
masses, PSMBHs, include the supermassive black holes know to exist
at galactic centres.The PIMBHs and PSMBHs are, like all stars and planets,
electrically neutral and so experience only the attractive
gravitational force. The PSMBHs may be formed from lighter seeds
such as PIMBHs by merging and accretion.

\bigskip

\noindent
At the next and final higher mass scale are the PEMBHs with at least a trillion
solar masses and carrying unscreened like-sign charges of order
$\sim 10^{40}$ Coulombs, a charge which corresponds only to a $\sim 10^{-18}$ 
charge asymmetry. These electric charges are not compensated by
accreted halos of opposite charge, and hence experience long-range
electromagnetic repulsion. This is the charged dark matter which replaces
what was given the name dark energy immediately after accelerated cosmic expansion
was discovered.

\bigskip

\noindent
The purpose of the present paper is to show that this model of
dark matter and energy \cite{FramptonMPLA} survives the generalisation
to spinning black holes. We had expected this to be far more difficult
than it is, as we have shown rather simply by extending the Reissner-Nordstrom
metric to the Kerr-Newman metric.

\bigskip

\noindent
If our EAU model is correct, the present decade promises to
resolve the dark matter and energy issues. The dark energy will have a
dual description either as charged dark matter or as a cosmological
constant $\Lambda(t)$ with constant equation of state $\omega(t) = P(t)/\rho(t)
= - 1$ with high accuracy. The  cosmological "constant'' itself falls in the
future as $\Lambda(t) \propto t^{-2}$.

\bigskip

\noindent
This leads to quite a different prediction for the future fate of the universe
than in the conventional $\Lambda CDM$ model with constant $\Lambda$.
Instead of exponential superluminal expansion of the extroverse
\cite{FramptonIJMPA}, the growth is as a power and the extroverse
does not grow within a trillion years to a gargantuan size and,
because of this far gentler rate of growth, billions of other
galaxies remain within the visible universe forever.

\bigskip

\noindent
In the EAU model, we need not include dark energy as a
slice of the energy pie-chart which become simply 5\% baryonic
matter and 95\% dark matter.With irrational optimism, we may expect that
by the end of this decade conventional
wisdom will change to where the accelerated expansion is ascribed
to electromagnetic repulsion rather than an unspecified form of repulsive
gravity.

\bigskip

\noindent
Let us re-emphasize that the EAU-model is speculative and based on
assumptions which have already been entertained in the existing literature
but which have not been rigorously justified. This fact was
already expressed {\it ut supra}. Our aim in the EAU-model
was to construct a cosmological theory in which dark energy, as 
originally understood, is removed and replaced by charged dark matter.

\bigskip

\noindent
Before delving further into the r\^{o}le of the PEMBHs, let us first address the simpler
question of why are the PIMBHs electrically neutral.
like stars and galaxies, while the PEMBHs carry negative electric charge?
This is due to the fact that the PIMBHs are formed during the
first second after the Big Bang when the selective accretion of electrons
over protons is unavailable. By contrast the PEMBHs are formed much later,
in the dark ages after the creation of the Cosmic Microwave Background (CMB),
when the difference in thermal velocities of electrons and protons leads
to the mechanism for acquisition of a net negative electric charge.

\bigskip

\noindent
In the EAU model, the same-sign-charged PEMBHs are so massive,
at least a trillion solar masses, that they are not associated with a specific galaxy or
cluster but move each on their own under the Coulomb repulsion
by other PEMBHs. An electric charge asymmetry $\epsilon_Q\equiv (Q_{+} - Q_{-})(Q_{+} +  Q_{-})^{-1}$ is induced in the intergalactic and intracluster gas due to the remaining
protons. It is straightforward
to estimate this charge asymmetry as $\epsilon_Q \sim 10^{-18}$ which survives
in the intergalactic and intracluster gas due to the residual protons. This
charge-asymmetric gas is not in the PEMBH environment so there is no
opportunity to set up Debye screening of the PEMBH negative charge and
hence long-range inter-PEMBH electric repulsion exists.

\bigskip

\noindent
There may have developed an incorrect prejudice, based on the
observed electric neutrality of stars and planets, of galaxies and of the dark
matter inside galaxies and clusters, that any macroscopic astrophysical object,
including a PEMBH, must be electrically
neutral. If the EAU model is correct, this prejudice must be discarded.

\vspace{0.5in}

\begin{center}
\section*{Acknowledgement}
\end{center}

\noindent
We are grateful to an anonymous referee for providing a summary of past research.

\end{document}